\documentclass[aps,prb,reprint,superscriptaddress,floatfix]{revtex4-2}

\usepackage[T1]{fontenc}
\usepackage{graphicx}
\usepackage{amsmath,amssymb,amsfonts}
\usepackage{bm}

\begin{document}

\title{Mode-selective electron-phonon coupling drives charge density waves in the kagome metals YRu$_3$Si$_2$ and LaRu$_3$Si$_2$}

\author{Wenqian Tu}
\affiliation{Key Laboratory of Materials Physics, Institute of Solid State Physics, HFIPS, Chinese Academy of Sciences, Hefei 230031, China}
\affiliation{University of Science and Technology of China, Hefei 230026, China}

\author{Run Lv}
\affiliation{Key Laboratory of Materials Physics, Institute of Solid State Physics, HFIPS, Chinese Academy of Sciences, Hefei 230031, China}
\affiliation{University of Science and Technology of China, Hefei 230026, China}

\author{Dingfu Shao}
\affiliation{Key Laboratory of Materials Physics, Institute of Solid State Physics, HFIPS, Chinese Academy of Sciences, Hefei 230031, China}

\author{Yuping Sun}
\affiliation{High Magnetic Field Laboratory, HFIPS, Chinese Academy of Sciences, Hefei 230031, China}
\affiliation{Key Laboratory of Materials Physics, Institute of Solid State Physics, HFIPS, Chinese Academy of Sciences, Hefei 230031, China}
\affiliation{Collaborative Innovation Center of Microstructures, Nanjing University, Nanjing 210093, China}

\author{Wenjian Lu}
\email{wjlu@issp.ac.cn}
\affiliation{Key Laboratory of Materials Physics, Institute of Solid State Physics, HFIPS, Chinese Academy of Sciences, Hefei 230031, China}

\date{\today}

\begin{abstract}
YRu$_3$Si$_2$ and LaRu$_3$Si$_2$ host the highest charge density wave (CDW) transition temperatures ever reported in kagome metals, $T_{\mathrm{CDW}}\approx 800$ and 400~K, with propagation vectors $Q_{\mathrm{CDW}}\approx(1/2,0,0)$ and $(1/4,0,0)$, respectively. The microscopic mechanism behind these record values has remained unresolved. Here, we combine anharmonic phonon renormalization extracted from molecular dynamics driven by a machine-learned force field with mode-resolved electron-phonon coupling (EPC) analysis to identify this mechanism. The CDW is driven by mode-selective EPC: the Ru out-of-plane phonon modes couple the Ru-$d_{xz}/d_{yz}$ and Si-$p_x/p_y$ orbitals, and the phonon linewidth peaks sharply at $Q_{\mathrm{CDW}}$, while the featureless electronic susceptibility rules out Fermi-surface nesting. Phonon anharmonicity melts the CDW, and the anharmonic phonon spectra reproduce both $Q_{\mathrm{CDW}}$ and $T_{\mathrm{CDW}}$ in good agreement with experiments. Molecular dynamics simulations also visualize the CDW melting in real space. Chemical bonding analysis further shows that the smaller Y$^{3+}$ radius strengthens the Ru-Ru bonds, enhancing lattice rigidity and accounting for the factor-of-two higher $T_{\mathrm{CDW}}$ of YRu$_3$Si$_2$. Our results establish a unified microscopic picture of CDW formation and melting in kagome metals YRu$_3$Si$_2$ and LaRu$_3$Si$_2$.
\end{abstract}



\maketitle

\section{Introduction}

Kagome lattice materials host Dirac cones, van Hove singularities (VHS), and flat bands arising from geometric frustration \cite{ref1,ref2,ref3}. The interplay of these features with electron-phonon coupling (EPC) gives rise to a rich phase diagram encompassing charge density waves (CDW) and superconductivity \cite{ref4}. Among kagome metals, the $A$V$_3$Sb$_5$ family ($A$ = K, Rb, Cs) has been most intensively studied, with CDW transition temperatures $T_{\mathrm{CDW}}\approx 80$--110~K \cite{ref5,ref6,ref7}. Higher $T_{\mathrm{CDW}}$ values have been reported in ScV$_6$Sn$_6$ ($\sim$90~K) \cite{ref8,ref9} and FeGe ($\sim$100--110~K) \cite{ref10}. The $R$Ru$_3$Si$_2$ ($R$ = Y, La) family, although first synthesized decades ago \cite{ref11}, has recently been found to host $T_{\mathrm{CDW}}$ far above these values. LaRu$_3$Si$_2$, a superconductor with a transition temperature $T_c\approx 7$~K, exhibits a primary CDW with $Q_{\mathrm{CDW}}\approx(1/4,0,0)$ at $T_{\mathrm{CDW}}\approx 400$~K and a secondary CDW with $Q_{\mathrm{CDW}}\approx(1/6,0,0)$ below $\sim$80~K, coexisting with magnetism and time-reversal symmetry breaking (TRSB) \cite{ref12,ref13,ref14}. Even more remarkably, YRu$_3$Si$_2$ exhibits a $T_{\mathrm{CDW}}\approx 800$~K, the highest ever reported in a kagome system, with $Q_{\mathrm{CDW}}\approx(1/2,0,0)$, alongside superconductivity below $T_c\approx 3.4$~K \cite{ref15,ref16}.

YRu$_3$Si$_2$ and LaRu$_3$Si$_2$ share the same high-temperature $P6/mmm$ parent phase, yet the microscopic mechanism of CDW formation in these record-high $T_{\mathrm{CDW}}$ kagome metals remains unresolved. A recent theoretical study emphasized the central role of EPC in LaRu$_3$Si$_2$ \cite{ref17}. However, that work was limited to the harmonic approximation, in which the $P6/mmm$ phonon spectrum displays extensive imaginary (unstable) modes. Because an imaginary soft mode leaves the phonon frequency and polarization vectors ill-defined, an EPC analysis performed directly on the harmonic spectrum is unphysical; the CDW instability must instead be studied after anharmonic renormalization removes these modes. Moreover, it remains open whether the same EPC mechanism extends to YRu$_3$Si$_2$, and what controls the factor-of-two difference in $T_{\mathrm{CDW}}$. Addressing these questions requires anharmonic lattice dynamics, as demonstrated for the prototypical kagome CDW systems CsV$_3$Sb$_5$ and ScV$_6$Sn$_6$ \cite{ref18,ref19,ref20}. Anharmonicity is thus a necessity, not a refinement, for these high-temperature CDW systems.

In this work, we combine anharmonic phonon renormalization extracted from molecular dynamics (MD) driven by a machine-learned force field (MLFF) \cite{ref21,ref22} and mode-resolved EPC analysis to elucidate the CDW mechanism in YRu$_3$Si$_2$ and LaRu$_3$Si$_2$. We identify mode-selective EPC, dominated by the Ru out-of-plane phonon modes that couple the Ru-$d_{xz}/d_{yz}$ and Si-$p_x/p_y$ orbitals, as the primary CDW driver, with the coupling strength peaking at the respective $Q_{\mathrm{CDW}}$. Our anharmonic phonon spectra reproduce both $Q_{\mathrm{CDW}}$ and $T_{\mathrm{CDW}}$ in good agreement with experiments and reveal that the CDW is melted by phonon anharmonicity, with MD simulations providing a direct real-space visualization of this melting. Chemical bonding analysis further reveals that the smaller Y$^{3+}$ ionic radius strengthens Ru-Ru bonds and thereby controls the CDW energy scale, accounting for the factor-of-two difference in $T_{\mathrm{CDW}}$. Together, our results offer a coherent microscopic account of how CDWs form and melt in kagome metals YRu$_3$Si$_2$ and LaRu$_3$Si$_2$.

\section{Methods}

First-principles calculations based on density functional theory (DFT) were performed using the Vienna \textit{ab initio} Simulation Package (VASP) \cite{ref23,ref24}. The generalized gradient approximation (GGA) in the scheme of Perdew-Burke-Ernzerhof (PBE) \cite{ref25} was applied for the exchange-correlation functional. A plane-wave energy cutoff of 500~eV and $\Gamma$-centered $k$-point meshes with a resolution of $0.02$~\AA$^{-1}$ for self-consistent calculations were used. The high-temperature $P6/mmm$ phases were fully relaxed until residual forces fell below $10^{-4}$~eV/\AA. For the harmonic phonon dispersion, we calculated harmonic force constants within a $4\times4\times4$ supercell using the finite difference method \cite{ref26} implemented in PHONOPY \cite{ref27}. Crystal orbital Hamilton population (COHP) analysis \cite{ref28} was performed using the LOBSTER package \cite{ref29}. The ``pbevaspfit2015'' basis set was adopted, including the $4p4d5s$ orbitals for Y, $5p5d6s$ orbitals for La, $4p4d5s$ orbitals for Ru, and $3s3p$ orbitals for Si.

We trained MLFFs for YRu$_3$Si$_2$ and LaRu$_3$Si$_2$ using an on-the-fly method during the DFT calculations, as implemented in VASP \cite{ref21}. The initial training datasets were generated via \textit{ab initio} molecular dynamics (AIMD) simulations at 100--900~K using a $4\times4\times4$ supercell to sample thermally perturbed ground-state configurations. AIMD simulations at the DFT level provide precise forces and energies for training. Because anharmonic phonon renormalization requires long trajectories at many temperatures, direct DFT MD is prohibitively expensive; the MD was instead driven by the MLFF using the LAMMPS package \cite{ref30}. The simulations were conducted in the canonical (NVT) ensemble, with the Nos\'e-Hoover thermostat applied to control the temperature \cite{ref31}. The renormalized phonon frequencies were extracted by projecting the MD trajectory onto a set of harmonic phonon modes using the mode decomposition technique as implemented in the DynaPhoPy code \cite{ref32}. For the real-space visualization of the CDW melting, the same MLFF and LAMMPS setup was used, but with the MD initialized from the experimentally reported CDW-distorted structures \cite{ref12,ref15} and a large $10\times10\times10$ supercell with 6000 atoms.

The EPC calculations were performed by combining the DFPT electron-phonon matrix elements with the anharmonic dynamical matrices following the approach of Ref.~\cite{ref33}. The electron-phonon matrix elements were first computed via DFPT \cite{ref34} using a modified version of Quantum ESPRESSO \cite{ref35} with projected augmented wave (PAW) pseudopotentials \cite{ref24}. A plane-wave energy cutoff of 60~Ry and a charge-density cutoff of 600~Ry were adopted. Electronic occupancies were treated using Gaussian smearing with a width of $\sigma=0.012$~Ry. The BZ was sampled with a $8\times8\times8$ mesh of $k$-points and a $4\times4\times4$ mesh of $q$-points. The anharmonically renormalized force constants obtained from DynaPhoPy were then converted to the dynamical matrix (.dyn) format and combined with the DFPT matrix elements to obtain the mode-resolved EPC constant $\lambda_{\mathbf{q}\nu}$, the Eliashberg spectral function $\alpha^2F(\omega)$, and the cumulative $\lambda(\omega)$. Furthermore, the electron linewidth $\Gamma_{n\mathbf{k}}$ was computed with the EPW code \cite{ref37}, which interpolates the DFPT electron-phonon matrix elements onto dense $k$-point grids via maximally localized Wannier functions obtained from Wannier90 \cite{ref38}. The phonon frequencies and polarization vectors entering the linewidth were obtained from the same anharmonically renormalized force constants and imported into EPW through its interface for reading renormalized interatomic force constants.

\section{Results and discussion}

YRu$_3$Si$_2$ and LaRu$_3$Si$_2$ crystallize in the hexagonal $P6/mmm$ (No.~191) space group in their high-temperature phase \cite{ref12,ref15,ref39}, with Ru atoms forming the kagome network in the $ab$-plane, sandwiched between Si layers and intercalated by Y/La cations (Fig.~1(a)). The primary structural difference originates from the smaller ionic radius of Y$^{3+}$ (0.90~\AA) versus La$^{3+}$ (1.03~\AA) \cite{ref40}: the DFT-relaxed lattice parameters are $a=5.71$~\AA\ and $c=3.57$~\AA\ for LaRu$_3$Si$_2$ (experimental $a=5.688$~\AA\ and $c=3.567$~\AA\ \cite{ref12}), and $a=5.53$~\AA\ and $c=3.66$~\AA\ for YRu$_3$Si$_2$ (no experimental $P6/mmm$ lattice constants are available, as this parent phase has not been directly observed \cite{ref15}). The in-plane lattice constant thus contracts by $\sim$3.2\% upon replacing La with Y, while the out-of-plane $c$ expands slightly. This in-plane contraction reduces the Ru-Ru bond length within the kagome layer from 2.86 to 2.77~\AA, with significant consequences for both the electronic structure and the chemical bonding.

\begin{figure*}[t]
\centering
\includegraphics[width=0.8\textwidth]{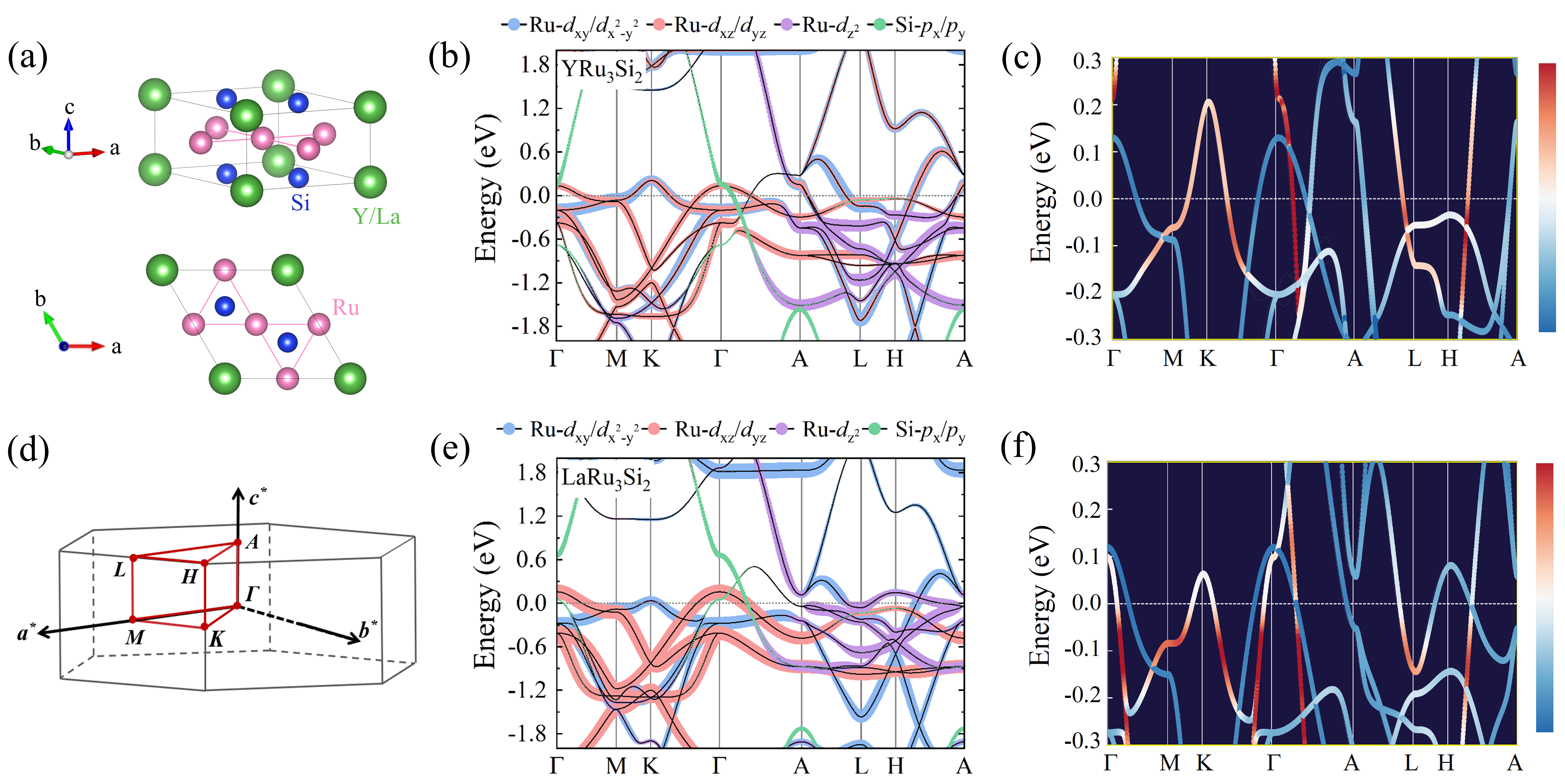}
\caption{(a) Side and top views of the YRu$_3$Si$_2$ and LaRu$_3$Si$_2$ crystal structures showing the kagome Ru layer (pink), Si layers (blue), and Y/La cations (green). (b) Orbital-projected band structure and (c) electron linewidth distribution of YRu$_3$Si$_2$. (d) Brillouin zone with high-symmetry points labeled. (e) Orbital-projected band structure and (f) electron linewidth distribution of LaRu$_3$Si$_2$.}
\label{fig:structures}
\end{figure*}

The orbital-projected band structures along the high-symmetry path $\Gamma$--$M$--$K$--$\Gamma$--$A$--$L$--$H$--$A$ (Fig.~1(d)) are shown in Figs.~1(b) and (e). Both compounds exhibit the hallmark kagome electronic features: Dirac-like crossings at $K$, flat bands along $\Gamma$--$M$ and $K$--$\Gamma$ arising from destructive interference of nearest-neighbor hopping on the kagome lattice, and VHS near the Fermi level $E_F$ associated with saddle points at the $M$ points \cite{ref15}. They also share the same dominant orbital character near $E_F$: the Ru-$d_{xz}/d_{yz}$ out-of-plane states (red) carry the largest spectral weight, and the $M$-point VHS provides a high density of these states at $E_F$, consistent with the experimental identification of two VHS near $E_F$ in YRu$_3$Si$_2$ \cite{ref15}. To establish which of these states couples most strongly to the lattice, we also calculate the electron linewidth, evaluated within the anharmonic EPC approach described in Sec.~II. The electron linewidth $\Gamma_{n\mathbf{k}} = 2\,\mathrm{Im}\,\Sigma_{n\mathbf{k}}$ shown in Figs.~1(c) and (f) is the electron-phonon scattering rate of each electronic state, with the imaginary part of the electron self-energy given in the Migdal approximation by \cite{ref41}:
\begin{multline}
\label{eq:linewidth}
\mathrm{Im}\,\Sigma_{n\mathbf{k}}(\omega) = \pi \sum_{m\nu} \int \frac{d\mathbf{q}}{\Omega_{\mathrm{BZ}}} \bigl|g_{mn,\nu}(\mathbf{k},\mathbf{q})\bigr|^2 \\
\times \bigl[ \delta(\omega - \varepsilon_{m,\mathbf{k}+\mathbf{q}} + \omega_{\mathbf{q}\nu}) + \delta(\omega - \varepsilon_{m,\mathbf{k}+\mathbf{q}} - \omega_{\mathbf{q}\nu}) \bigr]
\end{multline}
where $g_{mn,\nu}(\mathbf{k},\mathbf{q})$ is the electron-phonon matrix element, and $\omega_{\mathbf{q}\nu}$ is the anharmonically renormalized phonon frequency, with all energies measured relative to $E_F$. By comparing the $\Gamma_{n\mathbf{k}}$ distribution with the orbital-projected band structures, we identify the orbital character of the states that couple most strongly to the lattice. The largest electron linewidths occur on the Ru-$d_{xz}/d_{yz}$ and Si-$p_x/p_y$ derived bands, establishing these orbitals as the dominant electronic contributors to the EPC in both compounds.

To access the finite-temperature properties of the CDW beyond the static electronic structure, we turn to the anharmonic phonon calculations. Figures 2(a) and (c) display the harmonic (dashed line) and temperature-dependent anharmonic phonon spectra (colored solid lines) of the high-temperature $P6/mmm$ phase. The harmonic spectra are dynamically unstable across the entire acoustic branch along the $\Gamma$--$A$--$L$--$H$ plane ($Q_z=1/2$) for both compounds, with broad imaginary-frequency regions that obscure a specific soft-mode wavevector. In the anharmonic spectra, however, most of these unstable modes are stabilized by phonon-phonon interactions, and a well-defined soft mode emerges at $Q\approx(1/2,0,1/2)$ for YRu$_3$Si$_2$ (Fig.~2(a)) and $Q\approx(1/4,0,1/2)$ for LaRu$_3$Si$_2$ (Fig.~2(c)). These $q$-points correspond to the $L$ point and the midpoint of the $A$--$L$ line, respectively, in agreement with the experimentally observed CDW propagation vectors \cite{ref15,ref39}. Note that the standard experimental notation $(1/2,0,0)$ and $(1/4,0,0)$ refers to the orthorhombic $Cccm$ setting, whose $c$ axis is doubled relative to the $P6/mmm$ parent.

\begin{figure*}[t]
\centering
\includegraphics[width=0.8\textwidth]{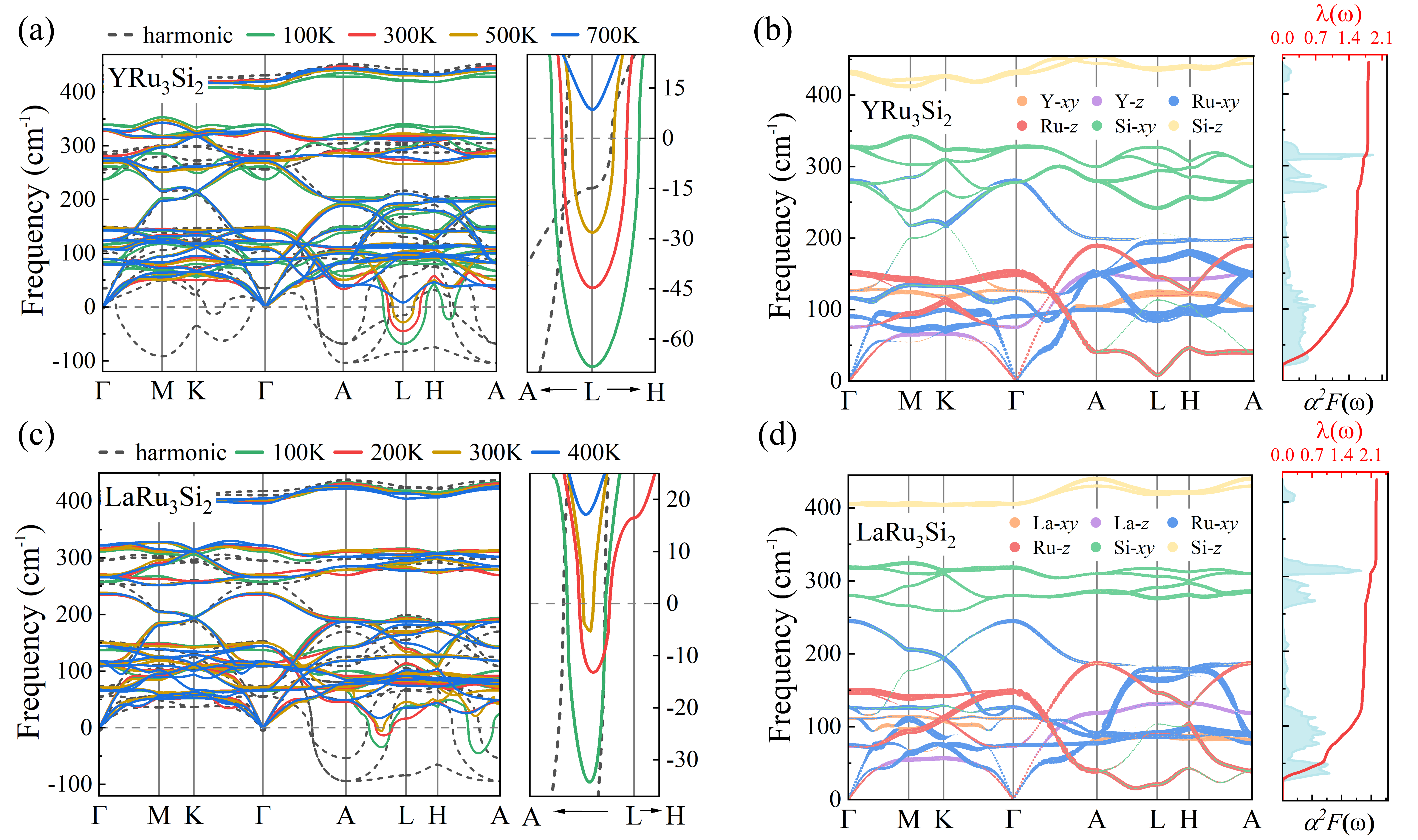}
\caption{(a) Harmonic phonon spectrum (dashed line) and temperature-dependent anharmonic phonon spectra (colored solid lines) for the high-temperature $P6/mmm$ phase of YRu$_3$Si$_2$. (b) Phonon dispersion of YRu$_3$Si$_2$ at 700~K, color-weighted by the contributions from Y, Ru, and Si atomic vibrations. The Eliashberg spectral function $\alpha^2F(\omega)$ (blue) and the integrated EPC constant $\lambda$ (red) are plotted on the right side. (c), (d) Same as (a), (b) for LaRu$_3$Si$_2$ at 400~K.}
\label{fig:phonons}
\end{figure*}

As temperature increases, anharmonic phonon-phonon scattering progressively renormalizes the soft mode upward. For YRu$_3$Si$_2$ (Fig.~2(a)), residual imaginary frequencies persist at 100, 300, and 500~K; the soft mode at $Q_{\mathrm{CDW}}\approx(1/2,0,1/2)$ is the last to stabilize, with the phonon spectrum achieving full stability only at 700~K. Throughout this work, $T_{\mathrm{CDW}}$ is defined by this dynamical criterion: the temperature at which the soft-mode frequency vanishes and the anharmonic phonon spectrum becomes fully real. This theoretical $T_{\mathrm{CDW}}$ underestimates the experimental value of $\sim$800~K \cite{ref15} by $\sim$12\%, which is comparable with the $\sim$15\% deviation observed in similar anharmonic calculations for CsV$_3$Sb$_5$ \cite{ref19}. For LaRu$_3$Si$_2$ (Fig.~2(c)), the spectrum becomes fully real at 400~K, consistent with its experimental $T_{\mathrm{CDW}}$ of $\sim$400~K \cite{ref39}. These spectra are free of any structural distortion; anharmonicity alone is responsible for stabilizing the high-symmetry phase. This stands in stark contrast to the harmonic picture, where imaginary frequencies imply a permanent instability. The result instead aligns with the established ``anharmonicity melts the CDW'' paradigm in which phonon-phonon scattering progressively stiffens the lattice until the soft-mode instability is overcome. While this mechanism has been demonstrated for CDW transitions below $\sim$100~K in CsV$_3$Sb$_5$ and ScV$_6$Sn$_6$ \cite{ref19,ref20}, YRu$_3$Si$_2$ and LaRu$_3$Si$_2$ extend its validity to $\sim$800~K and $\sim$400~K, respectively, far above that range.

We note that experimentally, both compounds exhibit an intermediate orthorhombic $Cccm$ phase between the $P6/mmm$ parent structure and the CDW-ordered $Pmma$ phase: $P6/mmm\rightarrow Cccm$ occurs at $\sim$880~K (YRu$_3$Si$_2$) and $\sim$600~K (LaRu$_3$Si$_2$), followed by the CDW transition at $\sim$800~K and $\sim$400~K, respectively \cite{ref15,ref39}. Our calculations start from the highest-symmetry $P6/mmm$ phase and neglect the intermediate $Cccm$ distortion. This simplification is well justified for LaRu$_3$Si$_2$, where the $\sim$200~K separation between the two transition temperatures implies that the CDW instability is only weakly affected by the prior $Cccm$ distortion. For YRu$_3$Si$_2$, however, the two transitions are separated by only $\sim$80~K; the proximity of these two instabilities suggests that the $Cccm$ distortion may non-negligibly renormalize the soft-mode frequencies, which could partially account for the $\sim$12\% underestimation of $T_{\mathrm{CDW}}$ in our $P6/mmm$-based calculation. A full treatment of the $Cccm$ intermediate phase lies beyond the scope of this work. Importantly, the $Cccm$ phase is itself dynamically stable: harmonic phonon calculations report no imaginary frequencies in the $Cccm$ structures of both YRu$_3$Si$_2$ \cite{ref15} and LaRu$_3$Si$_2$ \cite{ref17}, confirming the $Cccm$ phase as a dynamically stabilized distortion of the parent $P6/mmm$ phase.

Having identified the soft mode and its temperature evolution, we now examine the atomic character of the phonon modes. Figures 2(b) and (d) show the phonon dispersions of YRu$_3$Si$_2$ at 700~K and LaRu$_3$Si$_2$ at 400~K with atom-projected vibrational weights. These temperatures correspond to the full suppression of the imaginary frequencies. The low-frequency branches along $A$--$L$--$H$ are dominated by Ru out-of-plane character, with the soft CDW-active modes combining Ru out-of-plane and Si in-plane displacements. The atomic character of the soft modes identifies the degrees of freedom that participate in the distortion, and the anharmonic renormalization described above accounts for the temperature dependence of the soft mode and its melting. However, neither reveals why the lattice softens at $Q_{\mathrm{CDW}}$ in the first place. A possible mechanism is momentum-dependent EPC \cite{ref42}, which emphasizes the interaction between electrons and lattice vibrations through phonon-mediated inelastic scattering processes. To understand what drives this softening, we first compute the EPC at the temperatures where the anharmonic spectrum is fully real.

Within the anharmonic EPC approach described above, the electron-phonon matrix elements $\Delta_{ab}(\mathbf{q})$ are calculated to capture the deformation-potential coupling between electronic states at $E_F$ for each wavevector $\mathbf{q}$. These matrix elements are then combined with the anharmonic phonon frequencies $\omega_{\mathbf{q}\nu}$ and polarization vectors $e_{\mathbf{q}\nu}^a$ obtained from the DynaPhoPy calculation. This procedure replaces the unphysical harmonic frequencies (which are imaginary at $Q_{\mathrm{CDW}}$) with the anharmonically renormalized ones, relying on the established approximation that the deformation-potential matrix elements are dominated by the local electronic structure near $E_F$ and are insensitive to the moderate anharmonic renormalization of the phonon amplitudes. The Eliashberg spectral function $\alpha^2F(\omega)$ is then obtained from the anharmonic phonon frequencies and eigenvectors combined with the DFPT $\Delta_{ab}(\mathbf{q})$ matrix as:
\begin{multline}
\label{eq:eliashberg}
\alpha^2F(\omega) = \frac{1}{N_q} \sum_{abq\nu} \frac{e_{\mathbf{q}\nu}^a \cdot \Delta_{ab}(\mathbf{q}) \cdot e_{\mathbf{q}\nu}^{b*}}{2\,\omega_{\mathbf{q}\nu}\sqrt{m_a m_b}}\, \delta(\omega-\omega_{\mathbf{q}\nu})
\end{multline}
and the total EPC constant $\lambda$ is
\begin{equation}
\label{eq:lambda}
\lambda = \sum_{\mathbf{q}\nu}\lambda_{\mathbf{q}\nu} = 2 \int \frac{\alpha^2F(\omega)}{\omega}\,d\omega .
\end{equation}
Figures 2(b) and (d) also present the $\alpha^2F(\omega)$ (blue) and the cumulative $\lambda(\omega)$ (red). The cumulative EPC constant $\lambda$ reaches 1.8 for YRu$_3$Si$_2$ and 2.2 for LaRu$_3$Si$_2$, placing both compounds at the threshold of the strong-coupling regime. We stress that this BZ-averaged $\lambda$ does not by itself determine $T_{\mathrm{CDW}}$, which is instead set by the mode-selective EPC and the FS phase space at the soft-mode wavevector. The $\alpha^2F(\omega)$ spectra show that the dominant contributions to $\lambda$ come from low-frequency phonons, corresponding to the Ru-dominated acoustic and low-lying optical branches.

To resolve the momentum selectivity of this coupling, we calculate the phonon linewidth $\gamma(\mathbf{q})$, directly tied to the strength of the EPC. The wave-vector- and mode-resolved phonon linewidth $\gamma_{\mathbf{q}\nu}$ is defined as
\begin{multline}
\label{eq:phononlinewidth}
\gamma_{\mathbf{q}\nu} = 2\pi\omega_{\mathbf{q}\nu} \sum_{mn} \int \frac{d^{3}\mathbf{k}}{\Omega_{\mathrm{BZ}}} \bigl|g_{mn,\nu}(\mathbf{k},\mathbf{q})\bigr|^2 \\
\times \delta\bigl(\varepsilon_{n\mathbf{k}}-\varepsilon_F\bigr)\, \delta\bigl(\varepsilon_{m\mathbf{k}+\mathbf{q}}-\varepsilon_F\bigr)
\end{multline}
where $\omega_{\mathbf{q}\nu}$ is the frequency of phonon mode $\nu$ at wavevector $\mathbf{q}$, $g_{mn,\nu}(\mathbf{k},\mathbf{q})$ is the electron-phonon matrix element, and $\varepsilon_F$ is the Fermi energy. A peak in $\gamma(\mathbf{q})$ indicates strong coupling between phonons with wavevector $\mathbf{q}$ and electronic excitations near $E_F$. Figures 3(a) and (d) map $\gamma(\mathbf{q})$ in the $Q_z=1/2$ plane for YRu$_3$Si$_2$ at 700~K and LaRu$_3$Si$_2$ at 400~K, respectively, where $Q=(1/2,0,1/2)$ corresponds to the $L$ point and $Q=(1/4,0,1/2)$ lies midway along $A$--$L$. For YRu$_3$Si$_2$ (Fig.~3(a)), $\gamma(\mathbf{q})$ exhibits a sharp and isolated maximum precisely at $Q_{\mathrm{CDW}}=(1/2,0,1/2)$, with the intensity dropping by more than an order of magnitude away from this wavevector. The identical pattern is observed for LaRu$_3$Si$_2$ (Fig.~3(d)), where $\gamma(\mathbf{q})$ peaks sharply at $Q_{\mathrm{CDW}}=(1/4,0,1/2)$.

\begin{figure*}[t]
\centering
\includegraphics[width=0.7\textwidth]{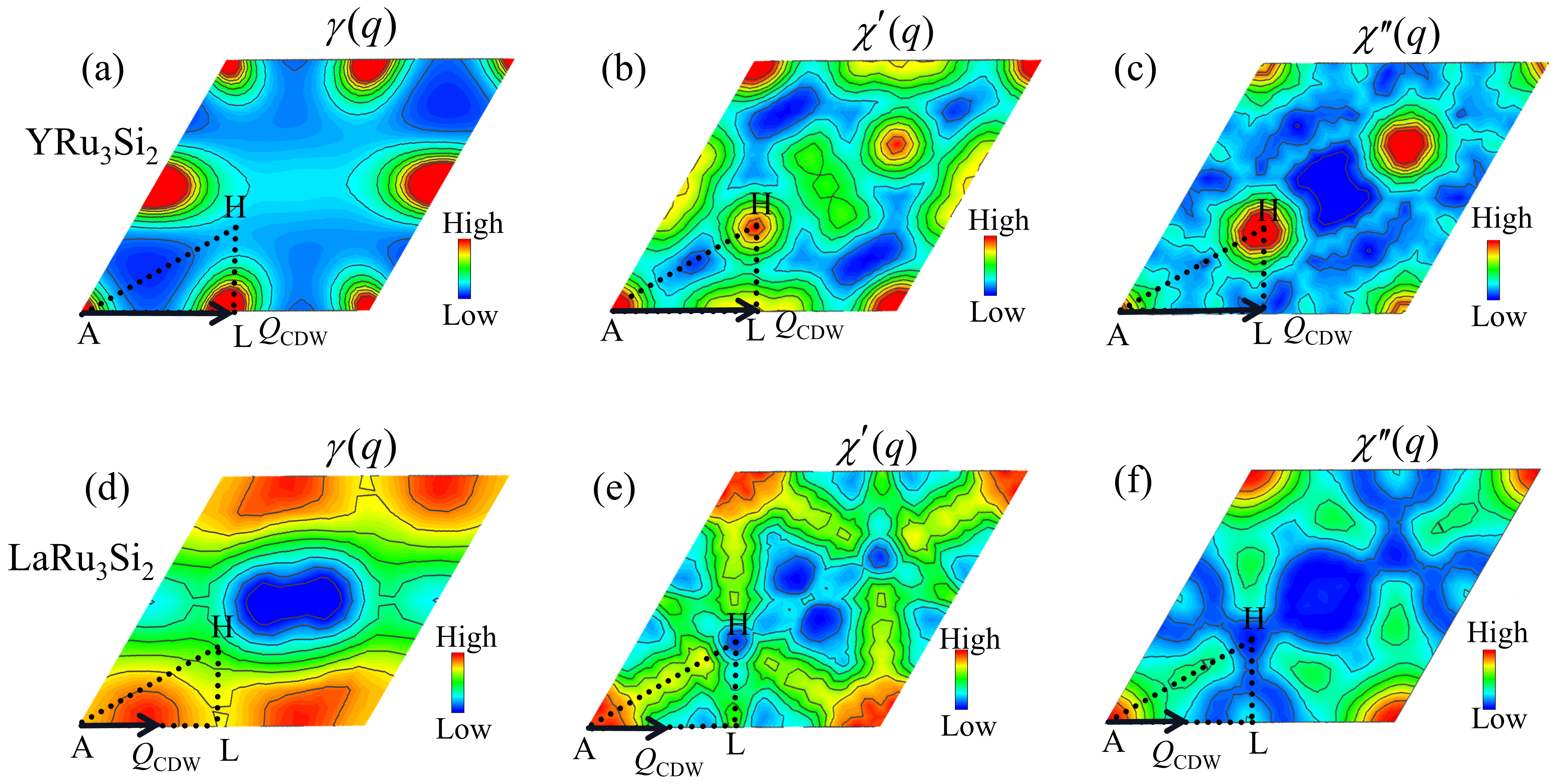}
\caption{(a) Phonon linewidth $\gamma(\mathbf{q})$ of the lowest phonon mode for YRu$_3$Si$_2$. (b) Real part $\chi'(\mathbf{q})$ and (c) imaginary part $\chi''(\mathbf{q})$ of the electronic susceptibility for YRu$_3$Si$_2$, where $Q_{\mathrm{CDW}}$ is indicated by a black arrow. (d)--(f) Corresponding results for LaRu$_3$Si$_2$. All panels display the quantities in the $Q_z=1/2$ plane.}
\label{fig:linewidth}
\end{figure*}

As established by the electron linewidth analysis, the electronic side of the coupling involves the Ru-$d_{xz}/d_{yz}$ and Si-$p_x/p_y$ orbitals, whereas the mode-resolved phonon analysis shows the CDW-active modes to be dominated by Ru out-of-plane displacements with a minor Si in-plane component. These two observations are reconciled by the mirror-symmetry selection rule of the $P6/mmm$ phase: the kagome Ru plane ($z=c/2$) and the Si/Y plane ($z=0$) are each invariant under the mirror $M_z$: $z\rightarrow-z$, which separates the orbitals into an $M_z$-even set ($\{d_{x^2-y^2},d_{xy},d_{z^2}\}$ and Si-$p_x/p_y$) and an $M_z$-odd set ($\{d_{yz},d_{xz}\}$ and Si-$p_z$) \cite{ref17}. Because the out-of-plane Ru displacement is $M_z$-odd, its matrix element $g_{mn}=\langle\psi_m|\partial V/\partial u_z|\psi_n\rangle$ vanishes unless $\psi_m$ and $\psi_n$ have opposite $M_z$ parity, so it can only scatter an $M_z$-even orbital into an $M_z$-odd one; among the states near $E_F$, the only inter-orbital channel accessible to these out-of-plane ($M_z$-odd) modes is Ru-$d_{xz}/d_{yz}$ (odd) $\leftrightarrow$ Si-$p_x/p_y$ (even). The coupling is thus mediated by the first-order modulation of the Ru-Si interlayer bond: because the Ru-Si bond spans the Si plane ($z=0$) to the Ru kagome plane ($z=c/2$), an out-of-plane Ru displacement stretches this Ru-Si interlayer bond to first order, whereas an in-plane displacement changes its length only to second order.

Another competing soft-phonon mechanism is Fermi surface nesting (FSN) \cite{ref43}, which attributes CDW formation to elastic electronic scattering at the FS, particularly in regions where parallel segments of the FS are separated by a single wave vector $Q$. To rule this out, we examine the real and imaginary parts of the electronic susceptibility $\chi(\mathbf{q})$ in the $Q_z=1/2$ plane. The real part $\chi'$ governs the stability of the electronic system, while the imaginary part $\chi''$ measures the phase space for low-energy electron-hole excitations. Therefore, if the CDW order is induced by FSN, both $\chi'$ and $\chi''$ should exhibit peaks at $Q_{\mathrm{CDW}}$. These parameters are defined by:
\begin{equation}
\label{eq:chi-real}
\chi'(\mathbf{q}) = \sum_{\mathbf{k}} \frac{f(\varepsilon_{\mathbf{k}})-f(\varepsilon_{\mathbf{k}+\mathbf{q}})}{\varepsilon_{\mathbf{k}}-\varepsilon_{\mathbf{k}+\mathbf{q}}}
\end{equation}
\begin{equation}
\label{eq:chi-imag}
\chi''(\mathbf{q}) = \sum_{\mathbf{k}} \delta\bigl(\varepsilon_{\mathbf{k}}-\varepsilon_F\bigr)\,\delta\bigl(\varepsilon_{\mathbf{k}+\mathbf{q}}-\varepsilon_F\bigr)
\end{equation}
Here $f(\varepsilon_{\mathbf{k}})$ is the Fermi-Dirac function, and $\varepsilon_{\mathbf{k}}$ is the corresponding Kohn-Sham energy. For YRu$_3$Si$_2$ (Figs.~3(b) and (c)), both the real and imaginary parts of the electron susceptibility exhibit modulations across the $Q_z=1/2$ plane, but their extrema do not coincide with $Q_{\mathrm{CDW}}=(1/2,0,1/2)$, ruling out any significant FSN contribution. The identical pattern is observed for LaRu$_3$Si$_2$ (Figs.~3(e) and (f)): the extrema of $\chi'$ and $\chi''$ do not align with $Q_{\mathrm{CDW}}=(1/4,0,1/2)$. This pronounced contrast between the featureless electronic susceptibility and the highly momentum-localized phonon linewidth provides direct evidence that the CDW in both compounds is EPC-driven rather than FSN-driven. The identical Ru out-of-plane character of the CDW-active phonons in both compounds, despite their different CDW periodicities, indicates that this coupling selectivity is an intrinsic geometric property of the Ru kagome lattice. Our results establish this out-of-plane mode-selective EPC mechanism across both LaRu$_3$Si$_2$ and the record-$T_{\mathrm{CDW}}$ compound YRu$_3$Si$_2$, demonstrating that the EPC-driven mechanism persists at temperatures far above those of any previously studied kagome CDW system.

\begin{figure*}[t]
\centering
\includegraphics[width=0.7\textwidth]{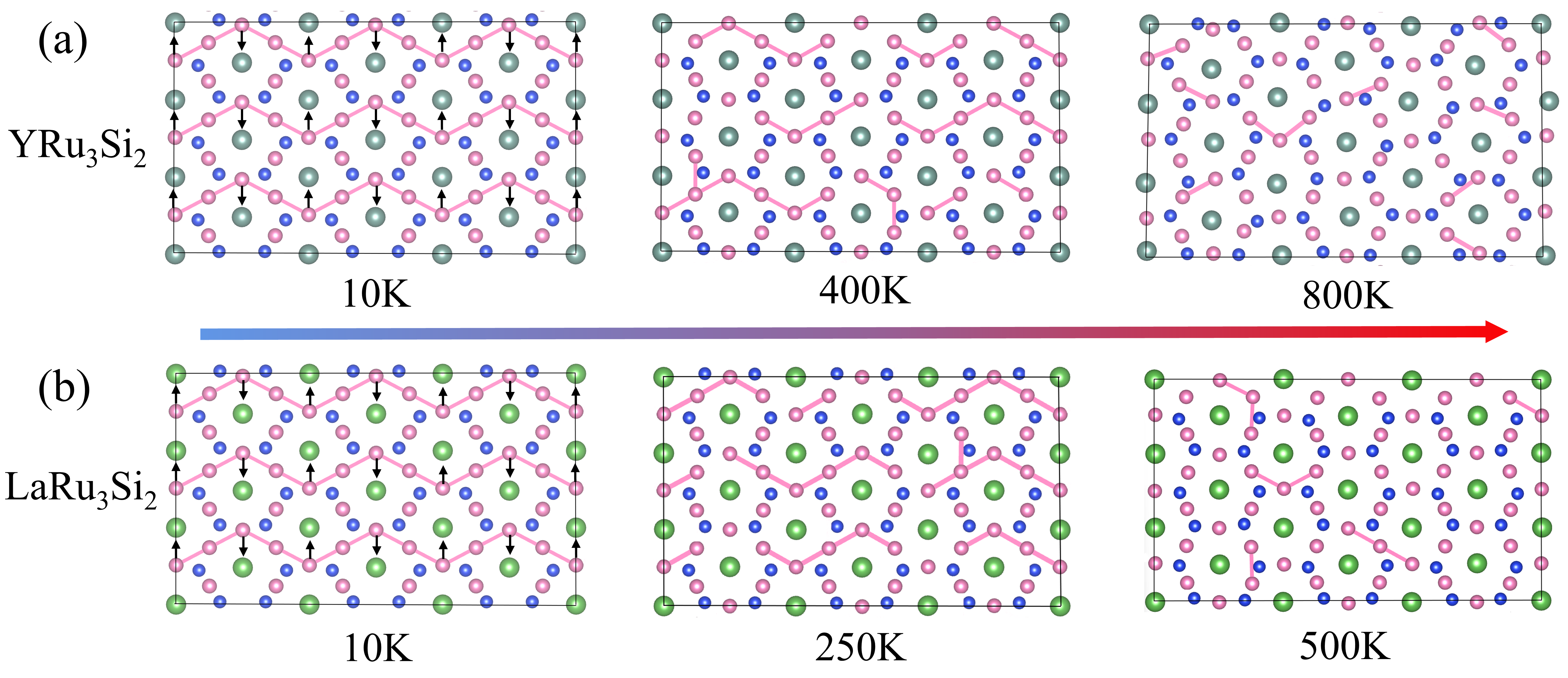}
\caption{MLFF-driven molecular dynamics simulations of the CDW melting process. (a) YRu$_3$Si$_2$ at 10~K, 400~K, and 800~K. (b) LaRu$_3$Si$_2$ at 10~K, 250~K, and 500~K. Black arrows indicate the Ru displacement directions relative to the high-symmetry positions, and pink lines mark the Ru-Ru bonds shortened by the CDW distortion, whose number progressively decreases with increasing temperature.}
\label{fig:md}
\end{figure*}

To directly visualize the CDW melting in real space, we performed MLFF-driven MD simulations starting from the experimentally reported CDW-distorted structures of YRu$_3$Si$_2$ \cite{ref15} and LaRu$_3$Si$_2$ \cite{ref12}. Figure 4 shows representative snapshots at 10~K, 400~K, and 800~K for YRu$_3$Si$_2$ (Fig.~4(a)) and at 10~K, 250~K, and 500~K for LaRu$_3$Si$_2$ (Fig.~4(b)), with the Ru displacement directions relative to the high-symmetry positions marked by black arrows and the shortened Ru-Ru bonds highlighted by pink lines. The experimentally reported CDW structures involve both in-plane and out-of-plane displacements of the Ru and Si atoms; to clearly visualize the distortion evolution, we display the in-plane component in Fig.~4. At 10~K, far below $T_{\mathrm{CDW}}$, the CDW-distorted Ru framework persists, with the Ru atoms displaced along the arrows from their high-symmetry positions. As the temperature rises toward and beyond $T_{\mathrm{CDW}}$, the distortion amplitude progressively diminishes. At sufficiently high temperatures, the Ru atoms relax back toward the undistorted high-symmetry equilibrium structure, with the shortened Ru-Ru bonds progressively disappearing, signaling the melting of the CDW. This real-space picture is consistent with the anharmonic phonon analysis above, in which phonon-phonon scattering progressively stiffens the soft mode until the high-symmetry phase is restored, and provides a direct atomic-level visualization of the ``anharmonicity melts the CDW'' mechanism.

\begin{figure*}[t]
\centering
\includegraphics[width=0.6\textwidth]{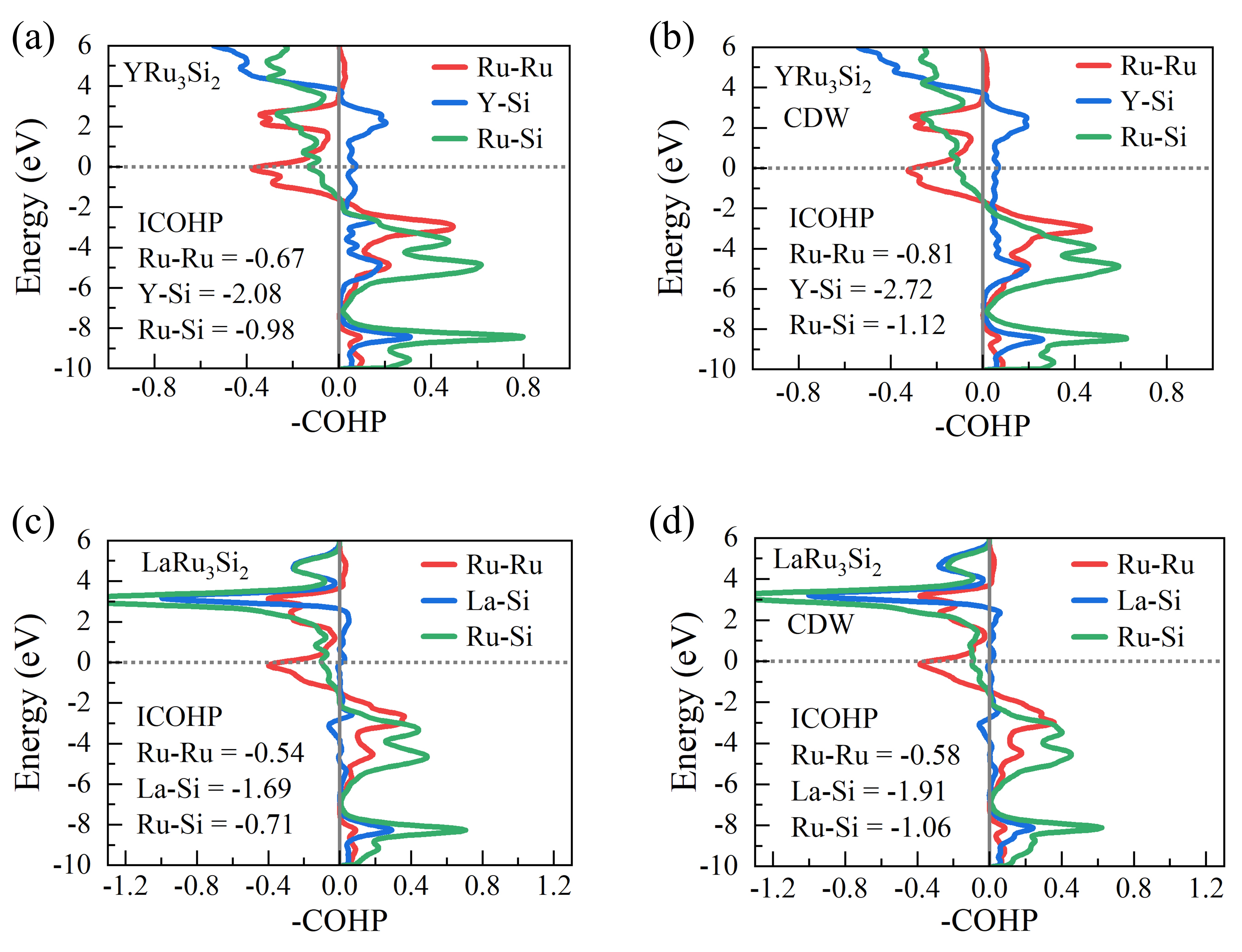}
\caption{COHP analysis of the parent $P6/mmm$ and CDW-distorted supercells. (a) YRu$_3$Si$_2$ parent, (b) YRu$_3$Si$_2$ CDW, (c) LaRu$_3$Si$_2$ parent, (d) LaRu$_3$Si$_2$ CDW. The curves show the COHP of the Ru-Ru, Y-Si (La-Si), and Ru-Si bonding pairs, and the numbers denote the corresponding ICOHP values.}
\label{fig:cohp}
\end{figure*}

The above results establish that EPC governs CDW formation in both compounds, but they do not explain the factor-of-two disparity in $T_{\mathrm{CDW}}$. To address this, we first evaluate the CDW formation energy, $\Delta E = E_{\mathrm{CDW}} - E_{\mathrm{normal}}$, defined as the total-energy difference between the CDW-distorted and undistorted supercells, normalized per formula unit. We obtain $\Delta E=-0.8$~eV per formula unit for YRu$_3$Si$_2$ and $-0.34$~eV per formula unit for LaRu$_3$Si$_2$, corresponding to an energy gain upon CDW formation that is $\sim$2.4 times larger in YRu$_3$Si$_2$. This larger thermodynamic driving force provides a natural explanation for the roughly twofold higher $T_{\mathrm{CDW}}$ of YRu$_3$Si$_2$ than that of LaRu$_3$Si$_2$. To identify the bonding origin of this energy-scale difference, we performed COHP analysis on both the parent and CDW-distorted supercells. The COHP methodology partitions the band-structure energy into pairwise orbital interactions, with negative (positive) values denoting bonding (antibonding) contributions; the integrated COHP up to $E_F$ (ICOHP) provides a quantitative measure of bond strength. Figure 5 presents the $-$\text{COHP} curves and ICOHP values for the dominant bonding pairs in all four structures. Comparing the parent phases first (Figs.~5(a) and (c)), all bonding pairs are systematically stronger in YRu$_3$Si$_2$ than in LaRu$_3$Si$_2$: for the Ru-Ru bonds that constitute the kagome network, the $-$\text{ICOHP} increases from 0.54~eV in LaRu$_3$Si$_2$ to 0.67~eV in YRu$_3$Si$_2$ (a 24\% enhancement), the Ru-Si bonds from 0.71 to 0.98~eV (38\%), and the Y/La-Si bonds from 1.69 to 2.08~eV (23\%). This is consistent with the geometric lattice contraction induced by the smaller Y$^{3+}$ radius, which shortens the Ru-Ru distance (2.77~\AA\ versus 2.86~\AA) and strengthens the covalent bonding. More importantly, the CDW distortion itself strengthens the bonding to a markedly greater degree in YRu$_3$Si$_2$ than in LaRu$_3$Si$_2$. Upon relaxation of the CDW-distorted Ru-Ru bonds, the structural element directly modulated by the CDW distortion gains 0.14~eV ($-$\text{ICOHP} from 0.67 to 0.81~eV, a 21\% increase) in YRu$_3$Si$_2$ (Figs.~5(a) and (b)), but only 0.04~eV (0.54 to 0.58~eV, 7\%) in LaRu$_3$Si$_2$ (Figs.~5(c) and (d)). The Ru-Ru bond strengthening in YRu$_3$Si$_2$ is more than three times larger than that in LaRu$_3$Si$_2$, paralleling its $\sim$2.4-fold larger CDW formation energy, indicating that the energy gain upon CDW formation is dominated by the strengthening of the Ru-Ru network.

These results establish a correlation between the Ru-Ru bond strength within the kagome network and $T_{\mathrm{CDW}}$. The stronger bonding in YRu$_3$Si$_2$ increases the lattice rigidity, which suppresses anharmonic thermal fluctuations that would otherwise melt the CDW, and consequently elevates the temperature at which anharmonicity overcomes the EPC-driven instability. The Y/La substitution thus acts as a chemical pressure that tunes the Ru-Ru bond strength without disrupting the kagome topology or the EPC mechanism. The EPC-driven CDW mechanism identified here is consistent with recent first-principles findings in other kagome metals: in CsV$_3$Sb$_5$, the CDW arises from large EPC and melts through lattice anharmonicity \cite{ref18,ref19}, and in ScV$_6$Sn$_6$, the competing CDWs are governed by anharmonic phonons \cite{ref20}. More importantly, our results reveal that the CDW energy scale can be tuned through chemical bond strength modulation via cation substitution, offering a practical route for engineering CDW properties in kagome metals.

\section{Conclusion}

In conclusion, we have presented a systematic study of CDW formation in YRu$_3$Si$_2$ and LaRu$_3$Si$_2$ combining anharmonic phonon renormalization extracted from MLFF-driven MD, mode-resolved EPC analysis, and quantitative chemical bonding characterization. We have shown that CDW formation in both compounds is governed by mode-selective EPC between the Ru-$d_{xz}/d_{yz}$ and Si-$p_x/p_y$ states, mediated by the Ru out-of-plane phonon modes. Phonon anharmonicity melts the CDW: the anharmonic phonon spectra reproduce both the experimental $Q_{\mathrm{CDW}}$ and predict $T_{\mathrm{CDW}}$ in good agreement with experiments, and MD simulations directly visualize this melting. Chemical bonding analysis traces the factor-of-two $T_{\mathrm{CDW}}$ disparity to bond strength: the smaller Y$^{3+}$ ionic radius strengthens the Ru-Ru network, yielding a stiffer lattice that suppresses the anharmonic fluctuations that would melt the CDW, elevating $T_{\mathrm{CDW}}$ to $\sim$800~K. These findings establish YRu$_3$Si$_2$ and LaRu$_3$Si$_2$ as model systems for EPC-driven CDW physics at extreme transition temperatures, and demonstrate that anharmonic lattice dynamics is essential for a quantitative understanding of CDW formation in kagome metals.

\begin{acknowledgments}
This work was supported by the National Key Research and Development Program of China under Contract No.~2022YFA1403203.
\end{acknowledgments}

\end{document}